\documentclass[a4paper,11pt]{article}
\usepackage{xcolor}
\pdfoutput=1 

\usepackage{jcappub} 

\usepackage[T1]{fontenc} 

\newcommand{\planck}{\textsc{Planck}}

\title{\bf Reheating after Starobinsky Inflation in the Jordan Frame}

\author{Gl\'auber C. Dorsch,}
\author{Luiz Miranda,}
\author{Nelson Yokomizo}

\emailAdd{glauber@fisica.ufmg.br}
\emailAdd{luizcarlos.miranda@hotmail.com}
\emailAdd{yokomizo@fisica.ufmg.br}

\affiliation{Departamento de F\'isica, Universidade Federal de Minas Gerais (UFMG),\\
Av. Ant\^onio Carlos 6627, Belo Horizonte, Minas Gerais, Brazil}

\abstract{We investigate gravitational reheating in the Starobinsky model in the Jordan frame, where inflation is driven by an $R^2$ modification of gravity with no explicit inflaton field. In this description, reheating proceeds exclusively through gravitational particle production triggered by the oscillations of the Ricci scalar after the end of inflation. We analyze the post-inflationary background evolution and show that an effective fluid emerging from the modified gravitational dynamics behaves as pressureless matter during the oscillatory phase. Including the backreaction of the produced particles, we demonstrate that the Ricci scalar oscillations acquire an exponential damping, consistently terminating particle production. Solving the coupled background and Boltzmann equations, we obtain a reheating temperature $T_{\mathrm{reh}} \sim 2 \times 10^{9}\,\mathrm{GeV}$. We finally compare the Jordan and Einstein frame descriptions and argue that, although classically equivalent, they can lead to distinct microphysical interpretations and quantitative predictions for reheating once quantum effects are taken into account.
} 

\begin{document}
\maketitle
\flushbottom

\section{Introduction}
\label{sec:intro}

Despite all the evidence corroborating the existence of an inflationary period at the onset of the cosmic history -- ranging from the CMB produced at very early times down to the large scale structure surveys of our cosmos today --, we still know very little about the dynamics that drives this cosmic inflation. The simplest mechanism involves postulating \emph{ad hoc} a new scalar field, dubbed the inflaton, whose potential energy dominates at the early Universe and leads to a fluid with negative pressure that drives the accelerated expansion. The shape of this inflaton potential is also established \emph{ad hoc}, but it is highly constrained by CMB data, such that a number of proposed single-field inflation scenarios can already be excluded. Alternatively, one can explore the possibility that cosmic inflation is the result of additional gravitational effects not contemplated by Einstein's theory of General Relativity, e.g. emerging from an action of the form
\[
    S = \int d^4 x \sqrt{-\det g_{\mu\nu}}~ f(R),
\]
with $f(R)$ some arbitrary function of the Ricci scalar $R$. Any non-linear function $f(R)$ leads to non-Einsteinian effects of gravity. Interestingly, one can always perform a field redefinition of the metric $g_{\mu\nu}$ to rewrite this action as a seemingly Einsteinian theory (i.e. linear in $R$) plus an effective scalar field, which emerges naturally from the modified gravity scenario, and can behave as the inflaton. This reparametrized theory is said to be in the \emph{Einstein frame}, whereas the original description in terms of a non-linear $f(R)$ (and in the absence of any inflaton field) is called the \emph{Jordan frame}.

At the classical level, the Einstein and Jordan frames are physically equivalent, as long as observables in one frame are translated into equivalent quantities in the other frame \cite{PhysRev.125.2163}. Whether this equivalence holds at the quantum level is an open question, with conflicting conclusions in the literature \cite{Flanagan:2004bz,Postma:2014,Kamenshchik:2014waa,Herrero-Valea:2016jzz,Pandey:2016unk}. In the context of cosmological applications, the descriptions of cosmic inflation in both frames were found to agree in a slow-roll approximation, under a map relating the slow-roll parameters and number of e-foldings in the distinct frames\footnote{Nevertheless, for a specified number of e-foldings $N_*$, the predictions of an inflationary model depend on which frame is taken to be the physical one, as this quantity is frame-dependent.}~\cite{Karam:2017zno}.

After inflation the Universe must undergo a reheating period, corresponding to the decay of the inflationary energy into what will eventually become radiation---thus kickstarting the $\Lambda$CDM evolution. For this process the equivalence of the two frames is even less evident. For instance, it has recently been shown that, in Starobinsky's theory of gravity, the modified gravitational interactions cause the inflationary energy to completely decay into radiation, leading to a successful reheating without additional \emph{ad hoc} assumptions~\cite{Dorsch:2024nan}.
This is clearly seen in the Einstein frame, where a trilinear coupling appears between the inflaton and other fields, leading to a decay channel that eventually depletes the inflaton's energy density and leaves only radiation.
 But in the Jordan frame there is no explicit inflaton field, hence such trilinear couplings responsible for the inflaton's decay are not manifest, and the physical interpretation of this reheating process is much more challenging. More generally, with the explicit inclusion of the quantum reheaton field, the equivalence between the two frames during reheating is not evident, and remains a topic of ongoing discussion~\cite{Faraoni:2006fx,faraoni1998conformaltransformationsclassicalgravitational,Belfiglio:2024swy}.

The purpose of this work is to perform a detailed analysis of the reheating process in the Jordan frame in a Starobinsky theory of gravity. We show that, as expected, the post-inflationary energy density is also depleted into radiation, but the analysis becomes somewhat more intricate. In the Jordan frame, inflation is caused by the modified Friedmann equations for the background metric. Reheating occurs due to the gravitational coupling of the background to other matter fields, which end up being excited at the end of inflation due to oscillations of the background, transferring energy from the background into radiation. Reheating can be completed because the backreaction of the produced radiation field eventually damps the background oscillations.

Before proceeding into the specificities, it is worth commenting on the status of the Starobinsky model when facing recently published data on the CMB spectrum. 
Starobinsky's model is an extremely attractive explanation for inflation precisely because it can fit well the CMB data with a minimal amount of fields and assumptions. Instead of having to introduce \emph{ad hoc} inflaton fields, with \emph{ad hoc} additional interactions to other fields to account for reheating, in this framework one minimally modifies the gravitational action by adding an $R^2$ term to the Einstein action, which is already expected from 1-loop quantum corrections to General Relativity~\cite{DeWitt:1967uc,Gurovich:1979xg,Vilenkin:1985md}. Surprisingly, this simple inflationary paradigm can successfully account for both inflation and post-inflationary reheating, being in excellent agreement with observations of the Cosmic Microwave Background (CMB) regarding the spectral index of scalar fluctuations as well as the bounds on the amount of primordial tensor fluctuations in the early Universe, e.g. as measured by the Planck collaboration~\cite{Planck:2018jri}. Recently other collaborations, such as the Atacama Cosmology Telescope (ACT) \cite{ACT:2025xdm,ACT:2025fju,ACT:2025tim} and the South Pole Telescope (SPT)~\cite{SPT-3G:2025bzu}, have published their analyses of the new CMB data which, combined with BAO data from DESI~\cite{DESI:2024uvr,DESI:2024mwx}, have pushed the spectral index of scalar fluctuations to slightly larger values than previous measurements, apparently putting the Starobinsky model into some tension with experiments~\cite{ACT:2025tim}. Three comments are however pertinent on this issue: (i) The new results reported by the ACT and SPT collaborations disfavour the Starobinsky model with a statistical significance of $\sim 2 \sigma$, which is not yet enough to rule out the model authoritatively;  (ii) the increased estimate for the scalar index was obtained through a combination of CMB data from ACT+Planck+SPT (called CMB-SPA) and BAO data from DESI, and these datasets are mutually inconsistent at $\gtrsim 3.1\sigma$~\cite{Ferreira:2025lrd,Kallosh:2025rni,McDonough:2025lzo} (therefore an even larger tension within the combined datasets than the $\sim 2\sigma$ tension between the estimated $n_s$ and the Starobinsky prediction). One should then be cautious against drawing strong conclusions before this BAO-CMB tension is itself resolved; (iii) the standard prediction for the spectral index in the Starobinsky model is a leading-order result that can be refined by accounting for higher-order corrections in the slow-roll expansion~\cite{Bianchi:2024qyp}. Furthermore, the precise value of $n_s$ depends on the number of e-folds $N_*$, which is intrinsically linked to the uncertainties in the duration and energy scale of the reheating epoch~\cite{Zharov:2025evb}. These refinements allow for a broader range of $n_s$ values that could potentially accommodate the recent ACT and SPT results. 

This paper is organized as follows. The theoretical framework of Starobinsky inflation is first established in section~\ref{2}, where we also compute the equations of motion for the background in the Jordan frame, setting the stage for the reheating analysis by defining the necessary initial conditions for the system's dynamics. In section~\ref{3} we perform a detailed investigation of the background evolution in the absence of backreaction effects. We provide fully numerical solutions for the equations of motion and derive analytical approximations that characterize the background behavior, while simultaneously exploring the underlying dynamics of the reheaton field during this phase. Our main finding in this section is that the inclusion of backreaction is essential for particle production to terminate. This we do in section~\ref{4}, finding that the backreaction leads to an exponential damping of the background oscillations, which in turn allows us to calculate the rate of gravitational particle production in the Jordan frame, from which we can determine the resulting reheating temperature of the Universe. In section \ref{5} we compare these results to those obtained in the Einstein frame. We discuss the subtle issue of frame inequivalence during the 
reheating period, highlighting how the physical interpretation may shift depending on the chosen representation. Finally, in section~\ref{6} we summarize our findings and present our concluding remarks on the implications of these results for Starobinsky's inflationary model.

\section{Evolution of the cosmic background in the Jordan frame}\label{2}

We are interested in studying the inflationary dynamics, and especially the post-inflationary reheating, in a Starobinsky framework of inflation, where the Einstein-Hilbert action is modified to
\begin{equation}
S = \frac{M_p^2}{2}\int d^4x \sqrt{-g} \left( R + \dfrac{R^2}{6M^2}\right)+ \int d^4x \mathcal{L}_m(g_{\mu\nu}, \chi) \; ,
\label{action}
\end{equation}
with $M_p=\sqrt{8 \pi G}$ the reduced Planck mass ($G$ being Newton's constant) and $M$ a constant with mass dimension that sets the scale of inflation. Einstein's General Relativity is obtained from the term $\sqrt{-g}R$ alone, so Starobinsky's quadratic term $ R^2$ introduces a modified gravity scenario, which suffices to explain inflation and the subsequent reheating process~\cite{Dorsch:2024nan,Jeong:2023zrv}. The lagrangian $\mathcal{L}_m$ describes the matter fields, which we will collectively label as a field $\chi$ dubbed the \emph{reheaton}, since it will describe the product of the reheating process that will eventually lead (through termalization or subsequent decay) to the radiation era of the $\Lambda$CDM. 

By varying the action with respect to the metric $g_{\mu\nu}$ we obtain the equation of motion
\begin{equation}
    R_{\mu\nu} - \frac{1}{2}g_{\mu\nu} R + \frac{1}{3M^2}\left(R_{\mu\nu} - \frac{1}{4}g_{\mu\nu}R - \nabla_\mu \nabla_\nu + g_{\mu\nu}\Box\right)R = \frac{1}{M_p^2}T_{\mu\nu},
    \label{EoMstarobinsky}
\end{equation}
where $T_{\mu\nu}$ is the energy-momentum tensor of the reheaton field given by
\begin{equation}
    T_{\mu\nu} = -\frac{2}{\sqrt{-g}} \frac{\delta \mathcal{L}_m}{\delta g^{\mu\nu}} \; .
\end{equation}

In an FLRW metric with scale factor $a(t)$, the Ricci scalar is
\begin{equation}
    R = 6\left(\frac{\ddot a}{a} + \frac{\dot a^2}{a^2}\right) = 6(2H^2 + \dot H),
    \qquad H\equiv \dfrac{\dot{a}}{a}=\text{Hubble parameter},
    \label{RfunctionofH}
\end{equation}
and the $t$-$t$  component of equation~\eqref{EoMstarobinsky} is given by
\begin{equation}
    \ddot H - \frac{\dot H^2}{2H} + \frac{M^2 H}{2} + 3H\dot H = \frac{M^2}{6HM_p ^2} \rho .
    \label{Hevolution}
\end{equation}
Equation~\eqref{Hevolution}, together with the definition of $H\equiv \dot{a}/a$, describes the evolution of $H$ and of the scale factor $a(t)$ in the \emph{Jordan frame} within the Starobinsky model, and consequently the evolution of $R$ through eq.~\eqref{RfunctionofH}.

It is also possible to derive an equation governing the evolution of the curvature scalar itself by taking the trace of eq.~\eqref{EoMstarobinsky}, which yields
\begin{equation}
    \ddot R + 3H\dot R + M^2\left(R + \frac{1}{M_p^2} T^{\ \mu}_\mu\right) = 0 .
    \label{Revolution}
\end{equation}
Equations~\eqref{Revolution} and \eqref{RfunctionofH} describe the evolution of $R$ in the Jordan frame. It is important to note that, in a vacuum, eq.~\eqref{Revolution} is identical to the Klein-Gordon equation for the inflaton in the Einstein frame. This implies that, in the Jordan frame, the curvature scalar behaves analogously to the inflaton. Therefore, during the reheating phase, we expect the oscillations of the Ricci scalar to be responsible for particle production, in analogy with the inflaton oscillations in the potential well in the Einstein frame.

Before proceeding to solve these equations for the cosmic background, let us note that eq.~\eqref{Hevolution} can also be rewritten in a more familiar form of a Friedmann equation, namely
\begin{equation}
    3 M_p^2 H^2 =  \rho + \rho_{\text{eff}} ,
    \label{rho_Rvsrho_eff}
\end{equation}
with $\rho_{\text{eff}}$ given by
\begin{equation}\begin{split}
    \rho_{\text{eff}}
    &= M_p^2\left( \frac{R^2- 12H\dot R}{12M^2 + 4R}\right).
    \label{rhoeff}
\end{split}\end{equation}
That is, even in vacuum $( \rho = 0)$, the background evolution involves an effective fluid. This fluid will play a crucial role during reheating because, in the absence of an inflaton in the Jordan frame, it must behave like matter during this phase. Reheating ends when the effects of this effective fluid become negligible compared to the energy density of the produced radiation. In the next section we will analyze the background evolution in vacuum, considering only the effects of the effective fluid.

\subsection{Initial conditions}
\label{Initialc}

Before solving the background equations we must determine the initial conditions for those differential equations. As in ref.~\cite{Dorsch:2024nan}, we will use \planck{} data to fix these initial conditions. During inflation, the first two terms on the left-hand side of eq.~\eqref{Hevolution} are negligible, and since no particles have been produced at this stage, we can write
\begin{equation}
    \frac{M^2}{2} = -3\dot H.
    \label{dotH}
\end{equation}
Integrating in time, taking as initial condition the instant $t_*$ where a certain pivot scale $k_*$ crosses the horizon, yields the behavior of \( H \) during inflation,
\begin{equation}
    H = H_* - \frac{M^2}{6}(t -t_*)\; .
    \label{Hinflation}
\end{equation}
Inserting this equation into the definition of the number of $e$-folds gives
\begin{equation}
    N \equiv \int^{t_f}_{t_i} H dt = H_* (t_f - t_*) - \frac{M^2}{12}(t_f - t_*)^2
    \label{efolds}
\end{equation}
where $t_f$ is the time at the end of inflation. To determine \( t_f \) we use the definition of the slow-roll parameter
\begin{equation}
\epsilon \equiv -\frac{\dot H}{H^2} \approx \frac{M^2}{6H^2} ,
\label{slow}
\end{equation}
noting that the end of inflation occurs at a time \( t_f \) when the slow-roll parameter satisfies \( \epsilon = 1 \). This gives
$    H_f = \frac{M}{\sqrt{6}}$,
which, plugging into eq.~\eqref{Hinflation} and solving for $t_f$, leads to
\begin{equation}
    t_f - t_* = \frac{6H_*}{M^2} - \frac{\sqrt{6}}{M} .
    \label{tf}
    \end{equation}
Substituting into into eq.~\eqref{efolds}, we find
\begin{equation}
    N_* \simeq  3\frac{H_*^2}{M^2}
    \label{efolds2}
\end{equation}
in regimes where $N \gg 1$. 
With eq.~\eqref{efolds2} we can finally determine the initial conditions using observational estimates of parameters of inflation from observations of the CMB . To do so, we will use the amplitude of scalar perturbations \( A_s \) and the number $N$ of e-folds. The scalar amplitude is related to background quantities in the Jordan frame through the following expression \cite{De_Felice_2010}:
\begin{equation}
    A_s = \frac{1}{12 \pi} \frac{M^2}{\epsilon^2}
        = \frac{N^2M^2}{3\pi},
    \label{Asm}
\end{equation}
where we used the well-known relation $N=1/2\epsilon$ between the slow-roll parameter \( \epsilon \) and the number of \( e \)-folds in the Jordan frame~\cite{De_Felice_2010,Bianchi:2024qyp}.

In the \planck{} papers, a pivot scale $k_*=0.05 \text{ Mpc}^{-1}$ is often chosen, and at this scale the observational constraint on $A_s$ determined from the \planck{} observations reads $A_s = 2.10(3) \times 10^{-9}$ \cite{Planck:2018jri}. However, in discussions related to inflation, an alternative choice of $k_*=0.002 \text{ Mpc}^{-1}$ is employed, which we also adopt here. Translating the scalar amplitude found by \planck{} for $k_*=0.05 \text{ Mpc}^{-1}$ to the alternative choice of pivot scale gives a best estimate of \cite{Bonga:2015xna,Bonga:2016iuf}
\begin{equation}
    A_s = 2.42 \times 10^{-9} \qquad (k_*=0.002 \text{ Mpc}^{-1}).
    \label{As2}
\end{equation}
In addition, \planck{} sets a constraint on the number of e-foldings for Starobinsky inflation, restricting it to the interval $49<N_*<59$ at a $95\%$ confidence level.

Therefore, using eq.~\eqref{As2} with \( N_* = 54 \), at the center of the allowed window (as in ref.~\cite{Dorsch:2024nan}), we can determine the value of \( M \) from eq.~\eqref{Asm}, and subsequently obtain \( H_\star \) from eq.~\eqref{efolds2}, thus fixing all initial conditions.


\section{Numerical evolution of the background without backreaction}\label{3}
\subsection{Numerical solutions}

At the start of the reheating process the Universe is still empty of matter and radiation, so the right-hand side of the Friedmann equation~ \eqref{Hevolution} vanishes. The evolution of the background is then determined by
\begin{equation}
    \ddot H - \frac{\dot H^2}{2H} + \frac{M^2 H}{2} + 3H\dot H = 0\; .
    \label{Hevolutionvacuum}
\end{equation}
To solve this equation, we need the initial values of \( M \), \( H_\star \), and \( \dot{H}_\star \). For this purpose, we use the results from the previous section. The values of \( M \) and \( H_\star \) for \( N = 54 \) are
\begin{equation}
    M = 3.36 \times 10^{13} \;\text{GeV}, \qquad H_*=1.42 \times 10^{14}\;\text{GeV}.
\end{equation}
For the initial value of \( \dot{H}_\star \) we use equation \eqref{dotH}, which shows that \( \dot{H}_\star \) is approximately constant during inflation. This is a good approximation, widely adopted in the literature\footnote{To obtain a more accurate value, one would need to solve the modified Friedmann equation without the approximation used in equation \eqref{dotH}, and consequently also use the value of another inflationary observational parameter, the scalar spectral index \( n_s \).}~\cite{Bianchi:2024qyp, De_Felice_2010}.

\begin{figure}
    \centering
    \includegraphics[scale=0.35]{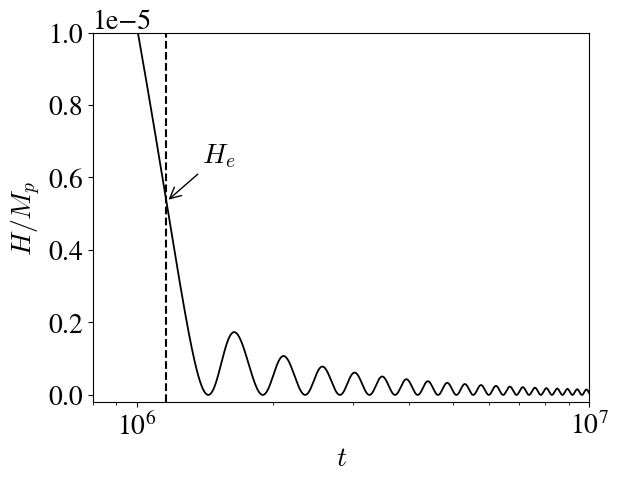}
    \includegraphics[scale=0.35]{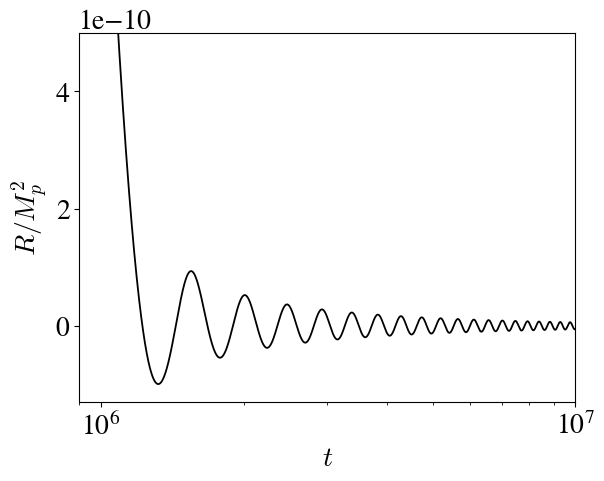}
    \caption{(Left) Hubble parameter as a function of proper time. The vertical dashed line represents the proper time when the first slow-roll parameter satisfies $\epsilon=1$, i.e. at the end of slow-roll inflation. (Right) Ricci scalar as a function of proper time.}
    \label{hubble_R}
\end{figure}

After solving eq.~\eqref{Hevolutionvacuum} with these initial conditions, we find the evolution of the Hubble parameter in proper time, shown in figure \ref{hubble_R} (left). From this solution, we can determine the value of the Hubble parameter at the end of inflation, that is, when the slow-roll parameter $\epsilon$ defined in equation~\eqref{slow} reaches unity,
\begin{equation}
    H_e = 1.34 \times 10^{13}\; \text{GeV}.
\end{equation}
The value obtained is slightly larger than the one found in the Einstein frame~\cite{Dorsch:2024nan}. This difference is expected, since in the Jordan frame the inflationary energy scale (which depends on \( H \)) is slightly higher due to the conformal transformation that affects the scale factor.

With the evolution of the Hubble parameter, we can also determine the evolution of the Ricci scalar using equation \eqref{RfunctionofH}, and the result is shown in figure~\ref{hubble_R} (right). As with the Hubble parameter, after the end of inflation the Ricci scalar starts to oscillate, and this oscillation will be crucial for particle production, serving as the necessary mechanism for reheating.

Figure \ref{hubble_R} reveals the first limitation of the numerical method: as reheating progresses, the oscillations of the Hubble parameter and the Ricci scalar become increasingly small, making the numerical integration more difficult.
\begin{figure}[h!]
    \centering
    \includegraphics[width=0.48\linewidth]{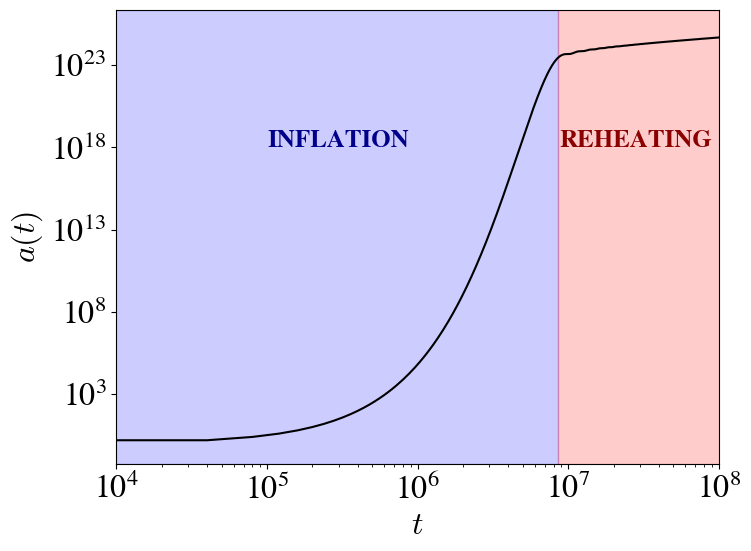}
    \quad
    \includegraphics[width=0.48\linewidth]{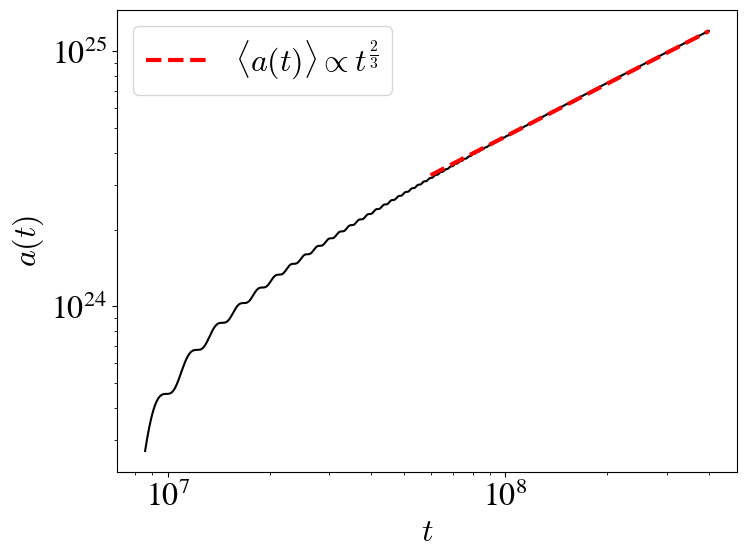}
    \caption{
        Evolution of the scale factor \( a(t) \) as a function of cosmic time \( t \) in log-log scale. ({Left}) Dynamics of the universe during inflation (blue region), characterized by an almost exponential expansion, followed by the reheating phase (red region). ({Right}) Zoom-in on the reheating phase, showing oscillations induced by the inflaton dynamics. The red dashed line corresponds to \(  a(t)  \propto t^{2/3} \), consistent with an effective equation of state \( w = 0 \), typical of a matter-dominated regime.}
    \label{fig:a_evolution}
\end{figure}
Since we have not yet considered the backreaction of particle production on the evolution of the background (because we are solving the Friedmann equation in empty space), at this stage the universe is dominated by the effective fluid described by eq.~\eqref{rhoeff}. The way this fluid evolves is crucial for understanding the reheating process and the parallel between the descriptions in the Einstein and Jordan frames.

In the Einstein frame, the oscillations of the inflaton around the minimum of the potential cause it to behave as matter with an equation of state $w=0$. In the Jordan frame there is no inflaton, but this same behaviour of a matter fluid can be  seen by studying the evolution of the scale factor, as shown in figure~\ref{fig:a_evolution}. The left panel shows the overall evolution of the scale factor during inflation and reheating. The right panel zooms-in on the reheating phase, and we see that after a sufficient number of oscillations the scale factor evolves as if the universe were matter-dominated, that is
\begin{equation}
    \langle a(t) \rangle\propto t^{\frac{2}{3}}\; .
    \label{adependence}
\end{equation}
Knowing that the universe is matter-dominated during reheating, we can say that the effective fluid governing the dynamics of the universe behaves as
\begin{equation}
    \langle \rho_{\text{eff}} \rangle = 3\langle H^2\rangle M_p^2\propto a^{-3}\; .
    \label{rhoeffdependence}
\end{equation}
So one sees that, just as in the Einstein frame, also in the Jordan frame the universe behaves as if matter-dominated during reheating, and later transitions to a radiation-dominated era through particle production. A necessary condition for the completion of reheating is that the radiation energy density \( \rho_R \) becomes larger than the effective fluid density \( \rho_{\text{eff}} \).

\subsection{Analytical approximations for the background evolution}
\label{sec:analytic_approx}

It is known that reheating after Starobinsky inflation lasts for approximately 20 e-folds in the Einstein frame~\cite{Dorsch:2024nan,Jeong:2023zrv}. Therefore, we expect the same to occur in the Jordan frame. However, solving the equations numerically throughout this entire period is challenging: the oscillations gradually reach extremely small amplitudes, requiring high numerical precision. To overcome this, it is useful to make analytical approximations for background quantities. To determine an analytical expression for the Hubble parameter, let us first neglect the damping term in equation~\eqref{Hevolutionvacuum}, rewriting it as
\begin{equation}
    \ddot H - \frac{\dot H^2}{2H} + \frac{M^2 H}{2} = 0 \quad\text{(neglecting damping)} .
    \label{Happrox}
\end{equation}
We can make the solution of Equation \eqref{Happrox} easier to visualize by performing a change of variable, \( H = u^2 \), showing that the undamped Friedmann equation behaves as an harmonic oscillator admitting an analytical solution
\begin{equation}
    H(t) \propto \cos^2\left(\frac{Mt}{2}\right)\; ,
    \label{Happrox2}
\end{equation}
with a constant multiplicative factor accounting for the amplitude of the oscillations. But in the numerical solution we saw that during reheating the amplitude of the oscillations in \( H \) \emph{decreases} over time. This is of course due to the damping term we neglected in the above analysis, and reinstating this term will lead to a time-dependent amplitude. Hence, inserting eq.~\eqref{Happrox2} with the time-dependent amplitude $\mathcal{A}(t)$ into eq.~\eqref{Hevolutionvacuum}, and assuming slow damping, one can find an expression for the amplitude~\cite{Mijic:1986iv}
\begin{equation}
    \mathcal{A}(t) = \frac{1}{3/M + (3/4)(t - t_{e}) + 3/(4M) \sin[M(t - t_{e})]}\; ,
    \label{At}
\end{equation}
where \( t_e \) is the cosmic time at the end of inflation.

An analytical expression for the Ricci scalar can then be found using equation~\eqref{RfunctionofH}. We are interested in using these analytical expressions at late times, after many oscillations that make the numerical analysis unfeasible. We can thus focus on the $M t \gg 1$ regime, where
\begin{equation}
R \simeq -\frac{4M}{t - t_{e}} \sin[M(t - t_{e})] \; .
\label{Rsolution}
\end{equation}
This was expected, since $R$ is described by equation~\eqref{Revolution}, which in vacuum is basically the equation of a harmonic oscillator damped by the expansion rate of the universe $H$, which decreases as $t^{-1}$.


\subsection{Reheaton dynamics}

Now let us look into the dynamics of the reheaton field to see how these particles are produced during reheating. In the Einstein frame this process is commonly described via an arbitrary coupling between the inflaton and other fields. However, in the Jordan frame there is no such thing as an inflaton field, and particle production is a purely gravitational effect from the coupling of all other fields to the background metric. 

For simplicity we will mimic the post-inflationary radiation content by a single massless scalar field $\chi$ dubbed the \emph{reheaton}. The action in the Jordan frame describing the dynamics of the massless reheaton field is
\begin{equation}
     S_\chi = -\int d^4 x \sqrt{-g}
        \frac12 g^{\mu\nu}\partial_\mu \chi \partial_\nu \chi,
        \label{reheatonaction}
\end{equation}
which yields the equation of motion
\begin{equation}
    \Box \chi = \dfrac{1}{\sqrt{-g}}\partial_\mu\left( \sqrt{-g} g^{\mu\nu}\partial_\nu \chi\right) = 0.
    \label{box_chi}
\end{equation}
In terms of the $k$-th mode $\chi_k$ of the field, this Klein-Gordon equation in an FLRW metric reduces to
\begin{equation}
    \ddot{\chi_k} + 3\frac{\dot{a}}{a} \dot{\chi_k} + \frac{k^2}{a^2}\chi_k = 0 .
    \label{21}
\end{equation}
In order to remove the friction term present in the equation, we introduce the conformal time $a d \eta = dt$ and the new rescaled field $\overline \chi = a \chi$. This leads to an equation of motion for this new field of the form
\begin{equation}
    \overline{\chi}_k^{\prime\prime}+ \omega_k ^2\overline{\chi}_k = 0 ,
    \label{modeequation}
\end{equation}
where $^\prime$ denotes derivative with respect to the conformal time and $\omega_k$ is the time-varying frequency, expressed by
\begin{equation}
    \omega_k(\eta) ^2 = k^2  -\frac{1}{6}a(\eta)^2 R(\eta) .
    \label{frequency}
\end{equation}

\begin{figure}[h!]
    \centering
    \includegraphics[width=0.48\linewidth]{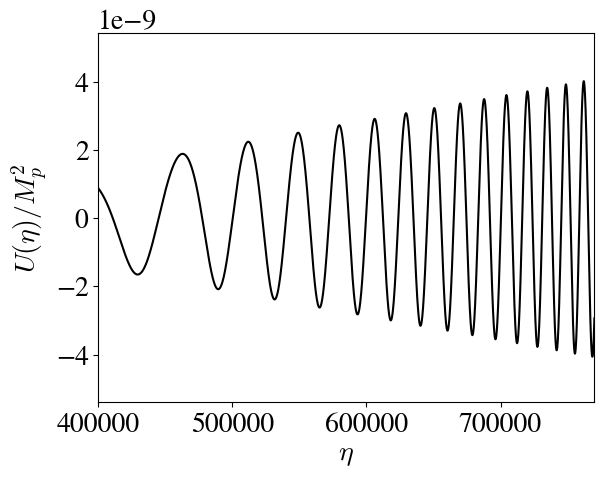}
    \caption{Behavior of the term responsible for particle production at the onset of reheating. The initial time corresponds to the first oscillation of the Ricci scalar.}
    \label{fig:U_evolution}
\end{figure}

Particle production is governed by the term
\begin{equation}
    U(\eta) = a(\eta)^2 R(\eta),
    \label{Ueta}
\end{equation}
since it is the only contribution responsible for non-adiabatic changes in the frequency.  The (conformal) time-dependence of this ``potential'' can be determined by solving the modified Friedmann equation \eqref{Hevolutionvacuum} in conformal time, 
\begin{equation}
    H'' = \frac{(H')^2}{2H} -\dfrac{M^2 a^2 H}{2} - 2aHH^\prime.
    \label{Hconformal}
\end{equation}
The result during the initial phase of reheating is shown in figure~\ref{fig:U_evolution}.

One sees that, when neglecting the backreaction effect of the reheatons and solving the background in a pure vacuum, the value of $U(\eta)$ oscillates with an ever increasing amplitude. This is not unexpected, since during reheating the universe is dominated by the effective fluid which behaves as matter, leading to a scale factor that evolves according to eq.~\eqref{adependence}, implying $a(\eta) \propto \eta^2$ in conformal time. Moreover, the Ricci scalar follows eq.~\eqref{Rsolution}, giving $R \propto \eta^{-3}$, so the term $U(\eta)$ evolves linearly with conformal time. 

However, this increasing behavior of $U(\eta)$ poses a physical problem because the density of particles produced at mode $k$ will be undamped even for very large modes, leading to an infinite number of total produced particles.  
This is an obvious indication that the effects of the created particles on the background evolution eventually become non-negligible and must be taken into account.

\section{Backreaction effects on background}\label{4}
In the previous section, we showed that particle production persists due to the growing amplitude of the oscillations in the term responsible for gravitational production during reheating. However, at some point this process must come to an end. To describe this regime, one must include the effect of backreaction by taking into account the right-hand side of eq.~\eqref{Hevolution} together with the energy-momentum tensor in eq.~\eqref{Revolution}. The inclusion of these effects, associated with reheaton production, leads to a damping of the Ricci scalar oscillations \cite{Arbuzova:2011fu, Dolgov:1998wz}, which causes the amplitude of $U(\eta)$ to decrease more rapidly in conformal time and, consequently, brings particle production to a halt. 

In this section, we follow the calculations presented in reference \cite{Arbuzova:2011fu}. To simplify the computations we work in conformal time, such that eq.~\eqref{Revolution} becomes
\begin{equation}
    R'' + 2\frac{a'}{a} + M^2 a^2 R = \frac{M^2}{a^2M_p ^2}T^\mu_\mu.
    \label{Revolutioneta}
\end{equation}

It is convenient to express the action \eqref{reheatonaction} in terms of conformal time $\eta$ and the rescaled field $\bar{\chi}=a\chi$,
\begin{equation}
    S_\chi = \frac{1}{2}\int d\eta\, d^3x\, \sqrt{-g}\,
    \left[
        (\bar{\chi}')^2 - (\vec{\nabla}\bar{\chi})^2
        + \frac{a^2 R}{6}\,\bar{\chi}^2
    \right].
    \label{actionconformal}
\end{equation}
From this action there follows the equation of motion for $\bar{\chi}$,
\begin{equation}
    \bar{\chi}'' - \nabla^2\bar{\chi} - \frac{1}{6}a^2 R\,\bar{\chi} = 0.
    \label{chisolution}
\end{equation}
The trace of the energy--momentum tensor of a massless scalar $\chi$ is given by
 $   T^{\mu}{}_{\mu} = g^{\mu\nu}\partial_\mu\chi\,\partial_\nu\chi  $
which, when rewritten in terms of the rescaled field $\bar{\chi}$ and plugged back into eq.~\eqref{Revolutioneta}, leads to
\begin{equation}
    R'' + 2\frac{a'}{a}R' + M^2 a^2 R = \frac{M^2}{a^2 M_{\!p}^2}\Big[(\bar{\chi}')^2 - (\nabla\bar{\chi})^2
      + \frac{(a')^2}{a^2}\,\bar{\chi}^2
      - \frac{a'}{a}\,(\bar{\chi}\,\bar{\chi}' + \bar{\chi}'\,\bar{\chi})\Big].
      \label{BR}
\end{equation}
Therefore, in order to determine the backreaction effects, we must evaluate the expectation value of the right-hand side of eq.~\eqref{BR}, namely \(\langle T^\mu_{\ \mu}\rangle\). To this end, we quantize the field \(\chi\) and obtain its evolution from eq.~\eqref{chisolution}. The solution of eq.~\eqref{chisolution} for \(R = 0\) is given by the free field, which can be expanded as~\cite{parker2009quantum}
\[
\chi^{(0)}(x) = \int\frac{d^3k}{(2\pi)^3}\frac{1}{2E_k}
\Big[\hat a_{\vec k}\,e^{-ik\cdot x} + \hat a_{\vec k}^\dagger\,e^{ik\cdot x}\Big],
\]
where \(x^\mu=(\eta,\vec x)\) and \(k^\mu=(E_k,\vec k)\) with \(E_k>0\). The creation and annihilation operators satisfy the usual commutation relation
\([ \hat a_{\vec k},\hat a_{\vec k'}^\dagger ]=(2\pi)^32E_k\delta^{(3)}(\vec k-\vec k')\).
The solution of eq.~\eqref{chisolution} including the gravitational coupling can be written in the integral form
\begin{equation}
    \chi(x) = \chi^{(0)}(x) - \frac{1}{6}\int d^4y\,G(x,y)\,a^2(y)\,R(y)\,\chi(y),
    \label{eq:sol}
\end{equation}
where the retarded Green's function of the wave operator reads
\begin{equation}
    G(x,y) = \frac{1}{4\pi r}\delta(\Delta\eta - r)\; .
\end{equation}
For a perturbative evaluation of the backreaction we expand the solution in powers of \(R\). 
At first order in \(R\), we replace \(\chi(y)\to\chi^{(0)}(y)\) inside the integral. For our purposes, we may assume that the gravitational coupling induces only a small perturbation on the free field, so it is sufficient to truncate the solution at first order in \(R\),
\begin{equation}
    \bar{\chi}(x) \approx \bar{\chi}^{(0)}(x) + \bar{\chi}^{(1)}(x),
\end{equation}
where
\begin{equation}
    \bar{\chi}^{(1)}(x) \equiv -\frac{1}{6}\int d^4y\,G(x,y)\,a^2(y)\,R(y)\,\chi^{(0)}(y).
\end{equation}
We then compute the expectation values using the free-field modes to obtain the contribution to the right-hand side of eq.~\eqref{BR}. 
The relevant terms are given by~\cite{Arbuzova:2011fu, Dolgov:1998wz}
\begin{align}
    \langle \chi^2 \rangle &\simeq -\frac{1}{48\pi^2} 
    \int_{\eta_0}^{\eta} d\eta' \frac{a^2(\eta')R(\eta')}{\eta - \eta'} 
    \label{eq:chi2}\\
    \langle \chi'^2 - (\vec{\nabla}\chi)^2 \rangle &\simeq 
    \frac{1}{96\pi^2} \int_{\eta_0}^{\eta} d\eta' 
    \frac{(a^2(\eta')R(\eta'))''}{\eta - \eta'}
    \label{eq:chiprime2}\\
    \langle \chi\chi' + \chi'\chi \rangle &\simeq 
    -\frac{1}{48\pi^2} \int_{\eta_0}^{\eta} d\eta' 
    \frac{(a^2(\eta')R(\eta'))'}{\eta - \eta'} 
    \label{eq:chichiprime}
\end{align}
The terms that depend only on $\chi^{(0)}$ do not contribute to particle production and have been neglected in the derivation leading to eqs.~\eqref{eq:chi2}--\eqref{eq:chichiprime}, since they are absorbed by renormalization.
From eq.~\eqref{Rsolution}, we know that $R' \sim M R$ and $R'' \sim M^2 R$. Since during reheating we are in the regime $M \gg H$, we have $Mt \gg 1$. Therefore, the term containing the second derivative in eq.~\eqref{eq:chiprime2} provides the dominant contribution to particle production. Consequently, eq.~\eqref{BR} can be written as
\begin{equation}
    R'' + 2\frac{a'}{a}R' + M^2 a^2 R \simeq
    \frac{M^2}{a^2 M_{p}^2}\frac{1}{96\pi^2}
    \int_{\eta_0}^{\eta} d\eta'
    \frac{(a^2(\eta')R(\eta'))''}{\eta - \eta'}.
    \label{Brequation}
\end{equation}

The most accurate way to analyze the backreaction effects on the background is to solve eq.~\eqref{Brequation} numerically. However, this is a rather difficult task, since during reheating the oscillations of the Ricci scalar become very small, which complicates the numerical integration.

To overcome this issue, we can make some approximations and attempt to find an analytical solution. Considering that the scale factor varies slowly and follows a power-law, $a(t) \propto t^p$, we found that the effective fluid dominating the Universe during reheating behaves like matter; hence, $p = \tfrac{2}{3}$. Therefore, we can factor out the scale factor from both the derivative and the integral.

Rewriting eq.~\eqref{Brequation} in terms of cosmological time, we obtain the integro-differential equation
\begin{equation}
\ddot R + 3H\dot R + M^2R \simeq
-\frac{1}{96\pi^2}\frac{M^2}{M_p^2}
\int_{t_0}^{t} dt' \frac{\ddot R(t')}{t - t'}.
\label{Ricci_backreaction}
\end{equation}
The right-hand side of equation~\eqref{Ricci_backreaction} is responsible for modifying the behavior of the Ricci scalar during the oscillatory phase, which causes the solution~\eqref{Rsolution} to become invalid in the presence of the reheaton. This is expected to induce a damping effect on the term \eqref{Ueta}, eventually causing particle creation to cease at some point.


\subsection{Gravitational production rate}\label{4.1}
In Section~\ref{sec:analytic_approx} we made some approximations to obtain analytical solutions for $H$ and $R$ in vacuum, given by Eqs.~\eqref{Happrox2} and~\eqref{Rsolution}. We can use these expressions to evaluate the effect of the backreaction on the amplitudes of $H$ and $R$.  

According to eq.~\eqref{Happrox2}, the asymptotic behavior of $H$ can be written as  
\begin{equation}
    H(t) = A(t) + B(t)\cos(\omega_rt)\; .
    \label{Hasym}
\end{equation}
The asymptotic behavior of $R$ is already given by eq.~\eqref{Rsolution}, but it is convenient to rewrite it as  
\begin{equation}
    R(t) = C(t)\sin(\omega_r t)\; ,
    \label{Rasym}
\end{equation}
where $A$, $B$ and $C$ are slowly varying functions of time and $\omega_r$ is not equal to M, given that the right side of equation \eqref{Ricci_backreaction} will give radiative contributions to the mass term $M$. 
Using the expression \eqref{Rasym}, we can determine the right-hand side of equation \eqref{Ricci_backreaction}:
\begin{equation}
\begin{split}
    g \int_{t_0}^{t} dt' \frac{\ddot R(t')}{t - t'} 
    &= g\cos(\omega_r t)\int_0^{t - t_0} \frac{d\tau}{\tau}\left[2\omega_r\dot C \cos (\omega_r \tau) - (\ddot C - \omega_r^2 C)\sin(\omega_r \tau)\right] \\
    &\quad + g\sin(\omega_r t)\int_0^{t - t_0} \frac{d\tau}{\tau}\left[2\omega_r\dot C \sin (\omega_r \tau) + (\ddot C - \omega_r^2 C)\cos(\omega_r \tau)\right]
\end{split}
\end{equation}
where
\begin{equation}
    g = -\frac{1}{96\pi^2}\frac{M^2}{M_p^2}, 
    \qquad 
    \tau = t - t'.
\end{equation} 

Now plugging \eqref{Rasym} and \eqref{Hasym} into \eqref{Ricci_backreaction} and comparing the coefficients of $\cos(\omega_r t)$ and of $\sin(\omega_r t)$ we obtain
\begin{equation}
    \ddot C + (M^2 - \omega_r^2)C + 3A\dot C 
    \simeq 
    g\int^{t - t_0}_{0} \frac{d\tau}{\tau}
    \left[
        2\omega_r \dot C \sin(\omega_r \tau) 
        + (\ddot C - \omega_r^2 C)\cos(\omega_r \tau)
    \right],
\label{A}
\end{equation}
\begin{equation}
    2\omega_r \dot C + 3\omega_r AC 
    = 
    g\int_0^{t - t_0} \frac{d\tau}{\tau}
    \left[
        2\omega_r \dot C \cos(\omega_r \tau)
        - (\ddot C - \omega_r^2 C)\sin(\omega_r \tau)
    \right].
\label{B}
\end{equation}
Note that, from Eqs.~\eqref{At} and \eqref{Rsolution}, we know that the amplitudes $A$, $B$, and $C$ scale as $1/t$. Consequently, we may neglect $\ddot C$ and $A\dot{C}$ compared to $C$ in eq.~\eqref{A}\footnote{This is also the reason why we could neglect terms involving $\cos^2(\omega_r t)$ and $\sin(\omega_r t)\cos(\omega_r t)$ since they are also accompanied by subdominant coefficients.}. Under these approximations, eq.~\eqref{A} yields the radiative correction to $M$,
\begin{equation}
    \omega_r^2 
    = 
    M^2 + gM^2 \int_0^{t - t_0} \frac{d\tau}{\tau}\cos(M\tau).
\end{equation}
Equation~\eqref{B}, in turn, gives the decay rate of the Ricci scalar amplitude. In the limit $t \gg M^{-1}$ the second integral on the right-hand side\footnote{The first integral, accompanying a $\dot{C}$ coefficient, will yield $\mathcal{O}(g^2)$ contributions which we neglect. The term $\ddot{C}$ is neglected for its subdominant $t^{-3}$ behaviour.} approaches $\pi/2$. By identifying $A(t) \simeq H$ as the non-oscillatory background component of the Hubble parameter we finally get
\begin{equation}
    \dot{C} + \frac{3}{2} H C = -\Gamma_G C + \mathcal{O}(g^2) \, ,
    \label{4.22}
\end{equation}
where we define the gravitational decay rate as
\begin{equation}
    \Gamma_G \equiv \frac{M^3}{384\pi M_p^2} \, .
    \label{rateJF}
\end{equation}
During the matter-dominated regime, $H=2/(3t)$, and the solution of this linear equation is
\begin{equation}
    C(t) = \frac{C_0}{t} e^{-\Gamma_G t} \, .
\end{equation}

Therefore, the asymptotic solution for the Ricci scalar undergoes an exponential decay due to the backreaction effect. Consequently, eq.~\eqref{Rsolution} should be corrected to
\begin{equation}
    R = -\frac{4M}{t - t_{e}} \sin\!\left[M(t - t_{e})\right] e^{-\Gamma_G (t- t_e)}.
    \label{Rsolution_back}
\end{equation}

This modified solution plays a crucial role in particle production, as it resolves the issue of the gravitational production term $U(\eta)$, which would otherwise grow indefinitely in the absence of backreaction. Thus, once backreaction is included, a characteristic timescale associated with $\Gamma_G$ emerges, causing the non-adiabatic variation in the reheaton field frequency to eventually cease.


\subsection{Reheating temperature}

Having determined the exponential damping of the Ricci scalar oscillations due to backreaction, we can now estimate the energy transferred to the fluid of gravitationally produced particles. The production rate obtained previously allows us to compute the energy density accumulated during the oscillatory phase. Once this fluid becomes relativistic, its energy can be associated with an effective temperature, which defines the gravitational reheating.

From the action~\eqref{actionconformal}, we see that the reheaton field is coupled to the gravitational sector through the term 
\(U = \frac{1}{6} a^2 R\). Therefore, the amplitude for the gravitational production of a pair of \(\chi\)-particles is
\begin{equation}
    \mathcal{M} \simeq \int d\eta\, d^3x\, 
    \frac{a^2 R}{6}\, 
    \langle p_1, p_2 | \chi \chi |0\rangle ,
\end{equation}
where
\begin{equation}
    |p_1, p_2\rangle 
    = \frac{1}{\sqrt{2}}\, 
      \hat a_{p_1}^\dagger 
      \hat a_{p_2}^\dagger |0\rangle .
\end{equation}

We define the gravitational source in conformal time as
\begin{equation}
    a^2(\eta) R(\eta) = \Omega(\eta)\, \sin(\omega \eta),
\end{equation}
where \(\Omega(\eta)\) varies slowly in conformal time and \(\omega = aM\) is the oscillation frequency. With this parametrization, we obtain
\begin{equation}
    |\mathcal{M}|^2 \simeq 
    \frac{\Omega(\eta)^2}{72}\,(2\pi)^4\,
    \delta^{(3)}(\mathbf{p}_1+\mathbf{p}_2)\,
    \delta(E_{p_1}+E_{p_2}-\omega)\,
    V\,\Delta\eta ,
    \label{11}
\end{equation}
where \(V\) and \(\Delta\eta\) denote the total volume and the conformal-time interval.

The particle production rate per unit volume and per unit conformal time follows from integrating~\eqref{11},
\begin{equation}
    n' = \int 
    \frac{d^3p_1\, d^3p_2}{4(2\pi)^6 E_{p_1} E_{p_2}}
    \frac{|\mathcal{M}|^2}{V\Delta\eta}
    \simeq \frac{\Omega(\eta)^2}{576\pi},
\end{equation}
in agreement with~\cite{Zeldovich:1977vgo, Vilenkin:1985md}.  
The gravitational rate of energy transfer into the \(\chi\) field is then
\begin{equation}
    \rho' = \frac{n'\,\omega}{2}
          = \frac{a M\, \Omega(\eta)^2}{1152\pi}.
\end{equation}

The production rate of the reheaton energy density is given by
\begin{equation}
    \dot{\rho}_\chi = \frac{M \langle R^2 \rangle}{1152\pi} ,
\end{equation}
where $\langle R^2 \rangle$ denotes the squared amplitude of the Ricci scalar oscillations. By employing eq.~\eqref{Rsolution_back}, the production rate can be expressed as a function of time during the reheating epoch:
\begin{equation}
    \dot{\rho}_\chi = \frac{M^3}{72\pi(t -t_e)^2}e^{-2\Gamma_G(t-t_e)} .
    \label{densityrate}
\end{equation}
Assuming that radiation is produced exclusively through the gravitational production of reheatons, the evolution of the radiation energy density, $\rho_R$, is governed by the Boltzmann equation with the source term defined in eq.~\eqref{densityrate}:
\begin{equation}
    \dot{\rho}_R + 4H\rho_R = \frac{M^3}{72\pi(t -t_e)^2}e^{-2\Gamma_G(t-t_e)} .
    \label{Boltzmann}
\end{equation}

The reheating temperature, $T_{\text{reh}}$, is defined at the moment when the radiation energy density coincides with the effective energy density $\rho_{\text{eff}}$, cf. eq.~\eqref{rhoeff}. From this point onward, the universe enters a radiation-dominated era. By numerically evolving the Boltzmann equation~\eqref{Boltzmann} alongside the effective energy density, and utilizing the averaged solutions for the Ricci scalar~\eqref{Ricci_backreaction} and the Hubble parameter~\eqref{box_chi}, we determine $T_{\text{reh}}$ through the relation:
\begin{equation}
    \rho_{\text{reh}} = \frac{\pi^2}{30} g_* T_{\text{reh}}^4 ,
    \label{stefan}
\end{equation}
where $g_* = 106.75$ represents the relativistic degrees of freedom. The initial values for the Hubble parameter and the Ricci scalar at the end of inflation are obtained from the numerical background solution.

\begin{figure}[ht]
    \centering
    \includegraphics[width=0.5\linewidth]{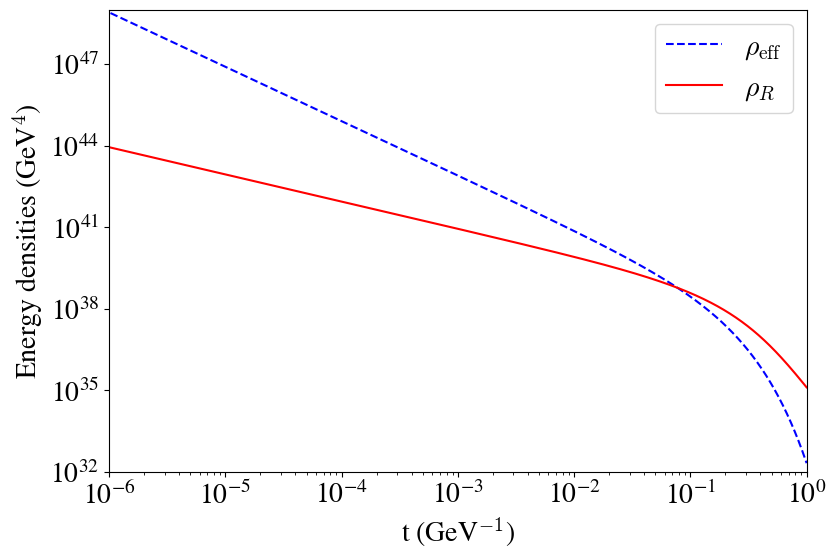}
    \caption{Evolution of the radiation and effective energy densities during reheating.}
    \label{fig:energy_densities}
\end{figure}

Figure~\ref{fig:energy_densities} illustrates the evolution of the energy densities over time. We observe that at $t_{\text{reh}} \approx 10^{-1} \, \text{GeV}^{-1}$, the densities equalize at a value of
\begin{equation}
    \rho_{R} = 5.5 \times 10^{38} \, \text{GeV}^4 .
\end{equation}
Consequently, substituting this result into eq.~\eqref{stefan}, the reheating temperature in the Starobinsky model within the Jordan frame is found to be
\begin{equation}
    T_{\text{reh}} = 2 \times 10^9 \, \text{GeV} .
    \label{TRJF}
\end{equation}

\section{Jordan vs. Einstein Frame}\label{5}

Having provided a comprehensive description of the reheating dynamics and particle production within the Jordan frame (JF), let us now establish its connection to the representation of the same process in the Einstein frame (EF).

\subsection{Einstein frame}

As demonstrated in \cite{Jeong:2023zrv,Dorsch:2024nan}, the phenomenology of reheating in the EF is typically interpreted through the decay of a scalar field evolving under an exponentially flat potential.
The transition to the Einstein frame is achieved by performing a conformal transformation on the metric, defined as
\begin{equation}
\tilde{g}_{\mu\nu} = \Omega^2(\phi) g_{\mu\nu} = e^{\sqrt{\frac{2}{3}} \frac{\phi}{M_P}} g_{\mu\nu} \, .
\label{conformaltransform}
\end{equation}
Under this transformation, the original action \eqref{action} is recast into the standard Einstein-Hilbert form plus a scalar field $\phi$ (the inflaton) with a canonical kinetic term. The dynamics are then dictated by the action
\begin{equation}
\tilde{S}_\phi = \int d^4 x \sqrt{-\tilde{g}} 
\left(
-\frac{M_P^2}{2} \tilde{R}
+ \frac{1}{2} \tilde{g}^{\mu\nu} \partial_\mu \phi \partial_\nu \phi - V(\phi)
\right) \, ,
\label{phiaction}
\end{equation}
where $V(\phi)$ is the Starobinsky inflaton potential, given by
\begin{equation}
V(\phi) = \frac{3 m^2 M_P^2}{4} \left( 1 - e^{-\sqrt{\frac{2}{3}} \frac{\phi}{M_P}} \right)^2 \, .
\label{potential}
\end{equation}

Furthermore, the conformal transformation significantly affects the reheaton action \eqref{reheatonaction}. During the reheating phase, an effective interaction term between the inflaton and the reheaton emerges \cite{Dorsch:2024nan},
\begin{equation}
    S_{\text{trilinear}} \approx \int d^4 x \sqrt{-\tilde{g}} \left( \frac{1}{M_P} \sqrt{\frac{2}{3}} \phi \, \tilde{g}^{\mu\nu} \partial_\mu \chi \partial_\nu \chi \right) \, .
\end{equation}
This trilinear coupling is responsible for reheaton production via inflaton decay. Using the standard Quantum Field Theory (QFT) formalism, the decay rate $\Gamma_E$ is found to be~\cite{Dorsch:2024nan}
\begin{equation}
    \Gamma_E = \frac{3 M^3}{1024\pi M_P^2},
    \label{rateEF}
\end{equation}
and consequently the reheating temperature in the EF is
\begin{equation}
    T_{\text{reh, EF}} \approx 4.5 \times 10^8 \; \text{GeV} \, .
    \label{TREF}
\end{equation}

\subsection{Frame inequivalence at reheating}

When analyzing the conformal factor \eqref{conformaltransform}, we observe that at the end of reheating, as $\phi \to 0$, the conformal factor approaches unity. One might therefore expect both frames to predict the same reheating temperature. However, comparing the reheating temperature obtained in the Jordan frame (JF) \eqref{TRJF} with that obtained in the Einstein frame (EF) \eqref{TREF}, we find that the temperature in the JF is somewhat higher than in the EF.

This discrepancy may indicate a lack of equivalence between the frames when quantum effects, such as particle production, are involved. Nevertheless, another possible explanation lies in the different approximations adopted in each frame.

\begin{figure}
    \centering
    \begin{picture}(200,130)
        \put(0,0){\includegraphics[width=0.5\linewidth]{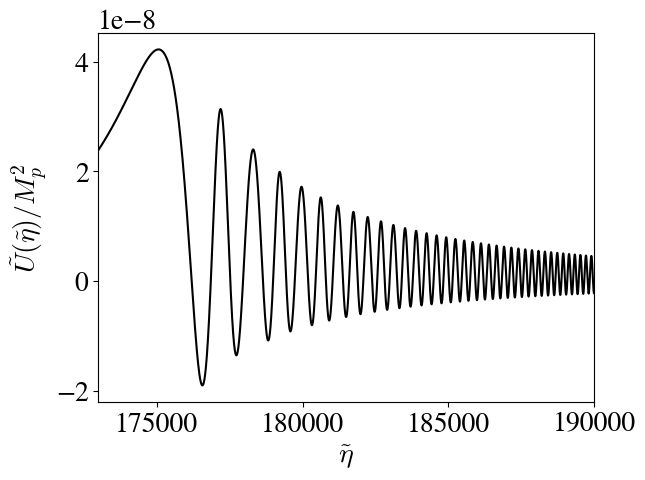}}
        \put(120,134){Einstein frame}
    \end{picture}
    \caption{Evolution of the source term \eqref{Ueta} in the Einstein frame. 
    The background equations in the Einstein frame were solved using the initial conditions of \cite{Dorsch:2024nan}. 
    The amplitude of the oscillations decreases with time, in sharp contrast to Fig.~\ref{fig:U_evolution}.}
    \label{fig:U_e}
\end{figure}
In the EF, particle production can be treated within two distinct regimes: a non-perturbative contribution driven by inflaton oscillations, and a perturbative contribution arising from the trilinear interaction between the inflaton and the reheaton field. This separation is possible because, in the EF, the oscillations of the source term \eqref{Ueta} rapidly become suppressed, as shown in Figure \ref{fig:U_e}. Consequently, once the non-perturbative production associated with oscillations becomes negligible, particle production proceeds only through the perturbative channel. However, the interaction action \eqref{reheatonaction} is treated approximately, and higher-order terms are neglected.

In contrast, in the JF there is no analogous separation: all particle production is driven by inflaton oscillations. Moreover, as illustrated in Figure \ref{Ueta}, the particle production source term does not decay with time. Nevertheless, as discussed in Section \ref{4.1}, once backreaction effects are included, particle production eventually ceases according to the gravitational production rate \eqref{rateJF}.

This difference may help explain why the production rate in the JF \eqref{rateJF} is more efficient than in the EF \eqref{rateEF}. In addition to corresponding to physically distinct mechanisms, the JF rate describes the entire production process, whereas the EF rate relies on approximations and does not fully account for the non-perturbative contribution.

\section{Conclusions}\label{6}

In this work we analyzed the viability of gravitational reheating in the Starobinsky model, as it is perceived in the Jordan frame, where particle production is a purely gravitational effect driven by the oscillations of the Ricci scalar. By studying the background dynamics in the absence of particle production, we could show that an effective fluid dominates the Universe during inflation and reheating until radiation dominance is eventually achieved. On the other hand, our analysis of the particle production mechanism showed that accounting for the backreaction of the produced particles is essential for this process to terminate, since otherwise the production term would grow indefinitely over time, as illustrated in Fig.~\ref{fig:U_evolution}.

Indeed, upon inclusion of backreaction effects we found that the amplitude of the Ricci scalar oscillations is exponentially suppressed, consistent with previous studies~\cite{Arbuzova:2011fu, Arbuzov:2025wox}. By employing an analytical approximation for the Ricci scalar under the influence of backreaction, we determined the reheaton production rate during the reheating epoch, finding agreement with results in~\cite{De_Felice_2010, Arbuzova:2011fu}. Our findings indicate that, in the Jordan frame, the Universe enters a radiation-dominated era with a reheating temperature of $T_{\text{reh}} \approx 2 \times 10^9$ GeV. Consequently, we conclude that reheating in the Starobinsky model, as described in the Jordan frame, is both viable and efficient. 

{It is interesting to emphasize that a purely gravitational reheating is possible in the Starobinsky model only because of additional gravitational effects that naturally lead to a total depletion of the primordial inflationary energy, as shown here and in ref.~\cite{Dorsch:2024nan}. In General Relativity this would not work: even if an \emph{ad hoc} inflaton field would decay gravitationally into radiation, and even if this radiation eventually dominates at some point, the expansion of the Universe would dilute the radiation energy density faster than the inflaton's, at least as long as the latter behaves like matter at the end of reheating ($w_\text{reh}=0$), which is expected after the inflaton stops oscillating around the bottom of the potential well. For this reason, successful gravitational reheating in Einstein's gravity requires a stiff post-inflationary equation of state ($w_\text{reh}>1/3$). But this in turn results in a blue-tilted spectrum of primordial gravitational waves which would have already been detected~\cite{Haque:2022kez, Figueroa:2018twl}. Here, in the context of Starobinsky's inflation, we have instead shown that the inflationary energy density is naturally depleted, either by a trilinear coupling of the inflaton to the reheaton as seen in the Einstein frame~\cite{Dorsch:2024nan}, or due to the backreaction of the reheaton damping the background oscillations (cf. figure~\ref{fig:energy_densities}).} Notably, the effective fluid dominating this phase possesses an equation of state $w_{\text{reh}} = 0$, ensuring that this framework does not suffer from the overproduction of gravitational waves observed elsewhere.

In summary, we provided a comparative analysis between the Jordan and Einstein frames, focusing on the fundamental differences in their respective particle production mechanisms. Although these frames are classically equivalent, we argue that the microphysics of reheating reveals a significant departure when quantum effects are considered. Specifically, the geometric nature of energy transfer in the Jordan frame contrasts sharply with the scalar field decay mechanism in the Einstein frame. By highlighting these distinct physical pathways, we demonstrate that the particle production rate during reheating is frame-dependent, ultimately leading to divergent predictions for the final reheating temperature. This discrepancy underscores the ongoing debate regarding the quantum equivalence of frames and emphasizes the necessity of identifying the true physical representation of the early Universe.

\providecommand{\href}[2]{#2}\begingroup\raggedright\endgroup

\end{document}